\documentclass{elsart}
\usepackage{psfig}
\usepackage{natbib}

\def\Mdot{{\dot{M}}}
\def\Mbh{{M_{\rm BH}}}
\def\Msun{{M_\odot}}
\def\Rg{{R_{\rm g}}}
\def\tc{{\tau_{\rm c}}}
\def\lambdaie{{\lambda_{\rm ie}}}

\begin{document}
\runauthor{Kawaguchi, Shimura and Mineshige}
\begin{frontmatter}

\title{Spectral Energy Distributions of Active Galactic Nuclei 
from an Accretion Disk with Advective Coronal Flow}

\author[Kyoto,Gakushin]{Toshihiro Kawaguchi} 
\author[Yokohama]{Toshiya Shimura}
\author[Kyoto]{Shin Mineshige} 
\address[Kyoto]{Department of Astronomy, 
Kyoto University, Sakyo-ku, Kyoto 606-8502, Japan}
\address[Gakushin]{Research Fellow of the Japan Society for the Promotion of
Science}
\address[Yokohama]{International Graduate School of Social Sciences, 
Yokohama National University, Hodogaya-ku, Yokohama, 240-8501, Japan}

\begin{abstract}
To explain the broad-band spectral energy distributions (SED) 
of Seyfert nuclei and QSOs,
we study the emission spectrum emerging from a vertical 
disk-corona structure composed of a two-temperature plasma by solving 
hydrostatic equilibrium and radiative transfer self-consistently.
Our model can nicely reproduce the soft X-ray excess 
with $\alpha$ ($L_{\nu} \propto \nu^{-\alpha}$) 
of about 1.5 and the hard tail extending to 
$\sim$ 50~keV with $\alpha \! \sim \!$ 0.5.
The different spectral slopes ($\alpha \! \sim \!$ 1.5 
below 2~keV and $ \!\sim \!$ 0.5 above) are the results of 
different emission mechanisms:
unsaturated Comptonization in the former
and a combination of Comptonization, bremsstrahlung, 
and reflection of the coronal radiation at the disk-corona 
boundary in the latter.
\end{abstract}


\end{frontmatter}

\section{Introduction}

Recent multi-waveband observations of Active Galactic Nuclei (AGNs) have
established significant deviations in the spectral shape of big blue
bump from a blackbody one.
A number of authors have tried to distort the accretion-disk spectrum toward 
the high energy regime so that the disk can emit substantial soft 
X-ray radiation as is observed.
One promising idea is 
Comptonization within the disk in the vertical direction 
(e.g., [1, 2, 3]). 
The effect of Comptonization is more prominent at higher accretion rates.

However, there still remain discrepancies between
models and observations. i) Although Comptonization
tends to increase $\alpha$, the Far-UV (FUV) spectrum of 
a Comptonized accretion disk has $\alpha$ $\sim$ 1 at best ([1, 3]),
whereas observed FUV spectra of distant quasars 
seem to be steeper ($\alpha \sim$~1.8--2.2; [4]). 
ii) The observed spectral 
index in soft X-rays ($\alpha \sim 1.4$--$1.6$; 
 [5]) is not achieved by any disk models 
since they produce Wien bumps at high energy, 
thus exhibiting exponential roll-over. 
iii) Comptonized accretion disks cannot reproduce the hard
X-rays. 
Thus, hard X-ray emission should be treated as an additional 
component in these models.

We therefore propose a new model, aiming 
to produce the overall SED by a disk-corona model.
The observed composite spectrum 
is taken from [4, 5] (see [6] for issues to be kept in mind).

\section{Basic Assumptions}

The numerical code used in this study is basically the same as that of
[2, 3] except for some modifications.
A constant fraction, $f$, of mass accretion ($\Mdot$) is assumed to be
dissipated in the corona with a Thomson optical depth of $\tc$, where
we consider 
advective energy transport of protons as well as
radiative cooling of electrons (Figure 1). 
\begin{figure}[tb]
\centerline{\psfig{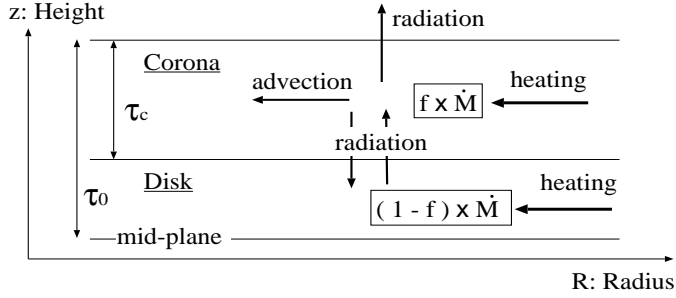}}
\caption{A schematic view of 
the configuration of our disk-corona model.}
\end{figure}
A remaining fraction, $1-f$, dissipates within the disk layer. 
The advective cooling rate in the corona 
is taken from the expression of an optically thin advection
dominated accretion flow (ADAF; [7])
The efficiency of the advection
is controlled by the viscosity parameter in the corona, $\alpha_c$.

The heating rate in the disk and corona, 
$q^+_{\rm d}$ and $q^+_{\rm c}$, are proportional to \hbox{$(1-f)\Mdot$} and 
$f\Mdot$, respectively, 
and the advective cooling rate in the corona per unit volume, 
$q^-_{\rm adv}$, is assumed to be 
proportional to the matter density.
The energy balance in each layer is as follows:
\vspace{-5mm}
\begin{eqnarray}
{\rm Disk:\ }   &{\rm Protons:\   }  q^+_{\rm d} = \lambdaie & \quad
                 {\rm Electrons:\ }  \lambdaie = q^-_{\rm rad} \\
{\rm Corona:\ } &{\rm Protons:\   }  q^+_{\rm c} = q^-_{\rm adv} + \lambdaie & \quad
                 {\rm Electrons:\ }  \lambdaie = q^-_{\rm rad},
\end{eqnarray}
where $\lambdaie$ is the energy exchange rate due to 
Coulomb collisions
and $q^-_{\rm rad}$ is the radiative cooling rate.
We consider free-free emission/absorption and  
Thomson/Compton scattering as radiation mechanisms.
The Comptonization is described by the Kompaneets equation. 

We divide the disk-corona from $300 \Rg$ to $3 \Rg$ into 20 rings 
so that each ring radiates approximately 
the same luminosity (cf. [1, 3]).
In total, the 
input parameters required for the calculations are
$M_{\rm BH}$, $\dot{M}$, 
$f$, $\tc$, $\tau_0$, and $\alpha_c$. 
The number of input 
parameters is similar to that of the relevant observed parameters 
which we aim to reproduce simultaneously; 
e.g., $L_X$, $\alpha_{\rm ox}$, $\alpha_{\rm UV}$, 
$\alpha_{\rm opt}$, $\alpha_{ROSAT}$, $\alpha_{ASCA}$. The 
spectrum of the whole disk-corona system is obtained by summing up the
emergent spectra of all the rings.

\section{Results}


We first show the most successful case with $M_{\rm BH} =3 \times 10^9 \Msun$
and $\dot{M} = 0.5 \ L_{\rm Edd} / c^2$
that corresponds to a luminosity of about 5\% of the Eddington luminosity, 
$L_{\rm Edd}$.
The thick line in Figure~2 shows an example of the resultant broad-band 
spectra.
A significant fraction $f = 0.6$ of mass accretion occurs via the
corona.
It turns out that the height of the disk-corona boundary 
measured from the mid-plane at $R$ = 5$\Rg$ is
0.03 $\Rg$, and that of the surface of the corona 
is 0.3 $\Rg$.
Then, the disk-corona system is indeed geometrically thin.

\begin{figure}[tb]
\centerline{\psfig{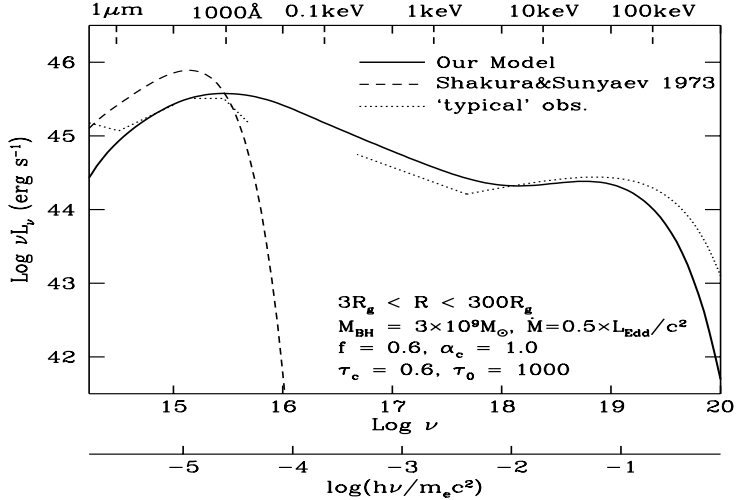}}
\caption{Resultant spectrum of the whole disk-corona integrated over 
3--300~$\Rg$ (thick line; from [8]).
Parameters used in this model are listed in the figure.
With those parameters, advective cooling in the corona is comparable 
to the radiative cooling at $R = 5 \Rg$; 
$q^-_{\rm adv} / q^+_{\rm c} \sim 0.5$.
The dashed line is the integrated spectrum of the standard disk 
with the same black-hole mass $\Mbh$ and $\Mdot$.
Dotted lines indicate the observed `typical' spectra which we aim to
fit with our models.
}
\end{figure}

The presence of multiple spectral components is the most noteworthy feature
of the present model. 
This is because different radiative mechanisms play roles 
in different wave-bands;
thermal radiation of the disk  in Opt./UV,
unsaturated Comptonization in FUV/soft X-rays, and a combination of 
unsaturated Comptonization, bremsstrahlung, and reflection in hard X-rays.
We wish to stress 
that the underlying radiative processes in soft--hard X-rays are
distinct from the traditional explanation, in which the UV--soft X-ray
component is due to blackbody radiation whereas the 
hard power-law component is due to Comptonization.
Note that the hard X-ray emission in our model looks only apparently 
like a power law with $\alpha \! \sim \!$ 0.5--1.0. 


The accretion-rate dependence of the emergent spectra
is shown in Figure 3.
As the accretion rate increases, the cut-off frequency of hard X-rays
(i.e., coronal electron temperature) decreases due to the increasing
efficiency of Compton cooling.
Then, the spectral slope at $\sim$ 0.03--1~keV 
gets steeper; $\alpha \!$ = 1.4, 1.7 and 2.1 
for $\Mdot / (L_{\rm Edd} / c^2) \!$ = 0.1, 0.5 and 0.7, respectively.
The qualitative trend that higher $\Mdot / L_{\rm Edd}$ 
leads to larger $\alpha_{ROSAT}$ is quite reminiscent of the cases of
Narrow-Line Seyfert 1s.
\begin{figure}[tb]
\centerline{\psfig{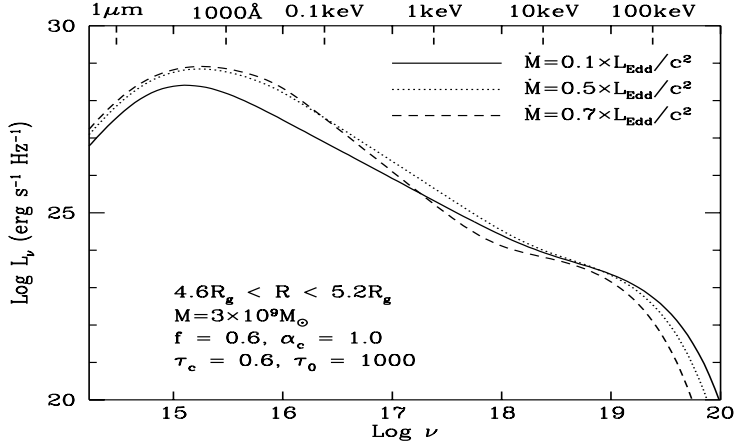}}
\caption{Accretion-rate dependence of the emergent spectra 
for a fixed black-hole mass ($\Mbh = 3 \times 10^9 \Msun$).}
\end{figure}
More detailed results and discussion will be presented elsewhere ([8]).


\end{document}